\documentclass[,final]{aipproc}

\layoutstyle{6x9}

\begin{document}

\title{A microscopic calculation of neutrino neutral reaction on $^{4}{\rm He}$}

\author{Doron Gazit}{
  address={The Racah Institute of Physics, The Hebrew University, 91904
Jerusalem, Israel.}
}

\author{Nir Barnea}{
  address={The Racah Institute of Physics, The Hebrew University, 91904
Jerusalem, Israel.}
}

\begin{abstract}
An ab-initio calculation of the inelastic neutral reaction of neutrino 
on $^{4}\mathrm{He}$ is presented \cite{OUR}, including realistic
nuclear 
force and full final state interaction among the four nucleons. The calculation uses the 
powerful combintation of the Lorentz integral
transform (LIT) method and the hyperspherical-harmonic effective interaction
approach (EIHH). The neutrino - nucleus interaction is taken in the impulse
approximation. With respect to previous calculations \cite{WO90}, \cite{HA88},
the current work predicts an increased reaction cross-section by 
$10\%-30\%$ for neutrino temperature up to $15$ MeV. 
\end{abstract}

\maketitle

Neutrino reactions with nuclear targets have an important part in many physical 
phenomena. In astrophysics, for example, neutrino-nuclei interactions have a main 
role in supernova explosion and the nucleosynthesis of the elements. The neutral
inelastic reactions of $^{4}\mathrm{He}$ with $\nu _{x}(\overline{\nu _{x}})$ 
($x=e,\mu ,\tau $) have a part in these two phenomena. Core collapse 
supernovae are widely accepted to be a neutrino driven
explosion of a massive star. When the iron core of a massive star becomes
gravitationally unstable it collapses until short-range nuclear forces halt
the collapse and drive an outgoing shock through the outer layers of the
core and the inner envelope. However, the shock loses energy through
dissociation of iron nuclei and neutrino radiation, and gradually stalls, it
becomes an accretion shock. It is believed, but to date not proven, that the
shock is then revived as neutrinos emitted from the collapsed core (the
proto-neutron star) deposit energy in the matter behind the 
shock to reverse the flow to an outgoing shock which explodes
the star. Hydrodynamic simulations fail in reviving the shock \cite{LI01},
\cite{BU03}. One of the ways to solve this problem is through enhancing the 
neutrinos-matter coupling. 

The matter behind the shock is composed
mainly of protons, neutrons, electrons, and $^{4}\mathrm{He}$ nuclei. In
contrast to the fairly known cross-sections of neutrinos with electrons and
nucleons, the interaction of neutrinos with $^{4}\mathrm{He}$ is not
accurately known. The current work is the first realistic microscopic 
calculation of this cross-section. 

In their way to the stalled shock, the electron-neutrinos remain in equilibrium
with matter for a longer period than their heavy-flavor counterparts, due to
the larger cross sections for scattering of electrons and because of charge
current reactions. Thus the heavy-flavor neutrinos decouple from deeper
within the star, where temperatures are higher. Typical calculations yield
temperatures of $\sim 10$MeV for $\mu $- and $\tau $- neutrinos
\cite{WI88}, which is approximately twice the temperature of
electron-neutrinos. Consequently, there is a considerable amount of 
$\nu _{\mu ,\tau }$ with energies above $20$ MeV that can dissociate 
the $^{4}\mathrm{He}$ through neutral reaction. 

Theoretical understanding of neutrino-nucleus scattering process is
achieved through perturbation theory of the weak
interaction model.
In the limit of small momentum transfer (compared to the Z particle rest
mass), the effective Hamiltonian can be written as a current-current interaction:
$\hat{H}_{W}=\frac{G}{\sqrt{2}}\int {d^{3}xj_{\mu }(\vec{x})J^{\mu }(\vec{x})}$,
where $G$ is the Fermi weak coupling constant, $j_{\mu }(\vec{x})$ is the
leptonic current, and $J^{\mu }$ is the hadronic current. The matrix element
of the leptonic current is $\langle f|j_{\mu }|i\rangle =l_{\mu }e^{-i\vec{q}
\cdot \vec{x}}$, where $\vec{q}$ is the momentum transfer and 
${l}_{\mu }=\bar{u}(k_{\nu ^{\prime }})\gamma _{\mu
}(1-\gamma _{5})u(k_{\nu })$.

The nuclear current matrix elements
consists of one body weak currents, but also many body corrections due to
meson exchange. The many-body currents are a result of meson
exchange between the nucleons. The current work is done in the
impulse approximation, thus taking into account only one-body terms. 
In order to estimate this approximation,
we refer to studies of inclusive electron scattering off 
$^{4}\mathrm{He}$ \cite{CA02}, where it is shown
that isovector electromagnetic two-body currents, which are proportional to the
electroweak vector currents, produce a strong enhancement of the transverse 
response at low and intermediate energies. In the current calculation, 
the vector part is almost negligible with respect to the axial part, and
the two-body axial currents are expected to give small contributions 
\cite{SC03}. 
The one-body currents connect the $^{4}\mathrm{He}$ 
ground state and final state wave functions. In order to calculate the
cross-section in a percentage level accuracy, one needs a solid
estimate of these wave functions.

\begin{table}[t] 
\begin{tabular}{cccc}
\hline
\hline
T [MeV] &  \multicolumn{2}{c}{$\langle \sigma \rangle_T$
[$10^{-42}cm^{2}$] }  &  \hspace{0.5cm} $\langle \sigma \omega \rangle_T$ \\
 & \hspace{0.15cm} This work & \hspace{0.25cm} Ref.~\cite{WO90} 
 & \hspace{0.5cm} $[10^{-40}cm^{2}\rm{MeV}] $\\ 
\hline
 4    &  2.09(-3) &    -     & 5.27(-4) \\
 6    &  3.84(-2) & 3.87(-2) & 1.03(-2) \\
 8    &  2.25(-1) & 2.14(-1) & 6.30(-2) \\
 10   &  7.85(-1) & 6.78(-1) & 2.30(-1) \\
 12   &  2.05     & 1.63     & 6.27(-1) \\
 14   &  4.45     &    -     & 1.42     \\
 16   &  8.52     &    -     & 2.84     \\
\hline
\hline
\end{tabular}
\caption{{\label{tab:crs}} Flavor and temperature averaged
inclusive inelastic cross-section and energy transfer cross-section
calculated. The temperatures are given in MeV, the cross-sections in
$10^{-42}cm^{2}$, and the energy transfer cross-sections in
$10^{-40}cm^{2}MeV$}
\end{table}

The differential cross-section is given by Fermi's golden rule, and is 
proportional to the response functions of the Coulomb, longitudinal, 
transverse electric and transverse magnetic multipole operators.
The response functions are calculated by combining two
powerful tools: the Lorentz integral transform (LIT) method
\cite{EF94} and the effective interaction hyperspherical
harmonics (EIHH) method \cite{BA00}. First we use the LIT
method in order to convert the scattering problem into a bound
state like problem, and then the EIHH method is used
to solve the resulting equations. Using this procedure we solve the
final state interaction problem avoiding continuum wave
functions.
The combination of the EIHH and LIT methods brings to a rapid
convergence in the response functions when increasing the effective interaction 
model space.

In Table~\ref{tab:crs} we present the calculated total temperature
averaged cross-section, $\langle\sigma\rangle_T=\frac{1}{2}
\frac{1}{A} \langle \sigma_\nu+\sigma_{\overline{\nu}}\rangle_T$, and
energy transfer cross-section, $\langle\sigma \omega \rangle_T=\frac{1}{2}
\frac{1}{A} \langle \omega \sigma_\nu+ \omega \sigma_{\overline{\nu}}
\rangle_T$, as a function of the neutrinos' temperature. Also
presented are earlier results by Woosley et. al. \cite{WO90}. It can
be seen that the current work predicts an enhancement of about
$10\%-30\%$ in the cross-section. 

The energy transfer cross-section was fitted by Haxton to the formula
\cite{HA88}, 
\begin{equation}
\langle\sigma \omega \rangle_T = \alpha \left( \frac{T-T_0}{10 \rm{MeV}}\right)^\beta
\end{equation}
with the parameters $\alpha=0.62 \cdot \rm{10^{-40} cm^2 MeV}$,
$T_0=2.54 \rm {MeV}$, $\beta=3.82$. A similar fit to our results
yields $\alpha=0.64 \cdot \rm{10^{-40} cm^2 MeV}$, $T_0= 2.05 \rm
{MeV}$, $\beta=4.46$. It can be seen that the current work predicts a stronger
temperature dependence of the cross sections. For example, a $15\%$ differnce 
between these calculations at $T=10$ Mev, grows to a $50\%$ difference at 
$T=16$ MeV.  

In conclusion, a detailed realistic calculation of the inelastic
neutrino-$^{4}\mathrm{He}$ neutral scattering cross-section is given. The
calculation was done in the impulse approximation with numerical
accuracy of about $1\%$. The different approximations used here should
result in about $10\%$ error, mainly due to many-body currents, which
were not considered in the current work.

The effect of these results on the supernova explosion mechanism
should be checked through hydrodynamic simulations, of various
progenitors. Nonetheless, it is
clear that our results facilitate a stronger neutrino-matter coupling
in the supernova environment. First, our calculations predict
an enhanced cross section by $10\%-30\%$ with respect to previous estimates.
Second, we obtained steeper dependence of the energy transfer cross-section
on the neutrino's temperature. Thus, supporting the observation
that the core temperature is a critical parameter in the
explosion process. 
It is important to notice that the energy-transfer due to inelastic
reactions are $1-2$ orders of magnitude larger than the elastic
reactions, ergo the inelastic cross-section are important to an
accurate description of the Helium shell temperature.

The Authors would like thank W. Leidemann, G. Orlandini, and L. Marcucci
for their help and advice. 
This work was supported by the {\bf ISRAEL SCIENCE
FOUNDATION} (grant no 202/02).

\bibliographystyle{aipproc}

\bibliography{nu_alpha_v2}

\end{document}